\documentclass[fleqn,10pt]{wlscirep}
\usepackage[utf8]{inputenc}
\usepackage[T1]{fontenc}
\usepackage{graphicx,subcaption}
\newcommand{\timestxt}{\!\times\!}

\title{Automatic airway segmentation from computed tomography using robust and efficient 3-D convolutional neural networks}

\author[1,2,*]{Antonio Garcia-Uceda}
\author[3,4]{Raghavendra Selvan}
\author[5]{Zaigham Saghir}
\author[1,2]{Harm A.W.M. Tiddens}
\author[1,3,+]{Marleen de Bruijne}
\affil[1]{Department of Radiology and Nuclear Medicine, Erasmus MC, 3015 CE, Rotterdam, The Netherlands}
\affil[2]{Department of Pediatric Pulmonology and Allergology, Erasmus MC-Sophia Children Hospital, 3015 CE, Rotterdam, The Netherlands}
\affil[3]{Department of Computer Science, University of Copenhagen, 2100, Copenhagen, Denmark}
\affil[4]{Department of Neuroscience, University of Copenhagen, 2200, Copenhagen, Denmark}
\affil[5]{Department of Medicine, Section of Pulmonary Medicine, Herlev-Gentofte Hospital, Copenhagen University Hospital, 2900, Hellerup, Denmark}
\affil[*]{a.garciauceda@erasmusmc.nl}
\affil[+]{marleen.debruijne@erasmusmc.nl}

\begin{abstract}
This paper presents a fully automatic and end-to-end optimised airway segmentation method for thoracic computed tomography, based on the U-Net architecture. We use a simple and low-memory 3D U-Net as backbone, which allows the method to process large 3D image patches, often comprising full lungs, in a single pass through the network. This makes the method simple, robust and efficient. We validated the proposed method on three datasets with very different characteristics and various airway abnormalities: i) a dataset of pediatric patients including subjects with cystic fibrosis, ii) a subset of the Danish Lung Cancer Screening Trial, including subjects with chronic obstructive pulmonary disease, and iii) the EXACT'09 public dataset. We compared our method with other state-of-the-art airway segmentation methods, including relevant learning-based methods in the literature evaluated on the EXACT'09 data. We show that our method can extract highly complete airway trees with few false positive errors, on scans from both healthy and diseased subjects, and also that the method generalizes well across different datasets. On the EXACT'09 test set, our method achieved the second highest sensitivity score among all methods that reported good specificity.
\end{abstract}

\begin{document}

\flushbottom
\maketitle
\thispagestyle{empty}


\section{Introduction}
\label{sec:Introduction}

Segmentation of the airway tree from thoracic Computed Tomography (CT) is a useful procedure to assess pulmonary diseases characterized by structural abnormalities of the airways, such as bronchiectasis (widening of the airways) and thickening of the airway wall. To quantitatively assess these conditions on the CT scan, clinicians are interested in having individual airway measurements, including airway diameter, wall thickness and tapering~\cite{Kuo2017,Kuo2020}. The bronchial tree is a complex 3D structure, with many branches of varying sizes and different orientations. Segmenting airways by manual or semi-automatic methods is extremely tedious and time-consuming, taking more than 15 h~\cite{Kuo2017} per scan, manually; or up to 2.5 h~\cite{Tschirren2009} per scan, semi-automatically. Thus, a fully automatic airway segmentation method is important to provide an accurate, effortless and free of user-bias segmentation of the airway tree needed to quantify airway abnormalities.

In CT scans, airways appear as tubular structures with an interior of typically low intensity (the lumen) enclosed by a structure of higher intensity (the airway wall). Moreover, airways are surrounded by a background which can be of low intensity (the lung parenchyma) or higher intensity (typically the vessels). The earliest airway segmentation methods used the region growing algorithm~\cite{Mori1996,Sonka1996} to segment the airway lumen, relying on the intensity difference between the airway lumen and wall. Region growing can accurately capture the central bronchi, but has a tendency to systematically fail at extracting the peripheral bronchi of smaller size, missing a large portion of the airway tree. Also, when segmenting the smaller airways it often results in segmentations leaking into the lung parenchyma. This is due to the reduced intensity difference and reduced signal-to-noise ratio between the airway lumen and wall, caused by partial volume effects near the smaller airways. Many airway segmentation methods build upon the region growing algorithm, using this as an initial step to segment the larger bronchi and then apply additional operators to detect smaller airways while preventing leakage~\cite{Fetita2004,Fabijanska2009,Graham2010,Lo2010,Lo2009}. Extensions to region growing based methods have been widely reviewed~\cite{Pu2012,Rikxoort2013}. In the airway extraction challenge EXACT'09~\cite{LoExact2012}, 15 algorithms were evaluated and compared, out of which 11 were region growing based methods. The results showed that all participating methods missed a significant amount of the smaller branches, and many methods had a large number of false positives errors.

Several airway segmentation methods use machine learning classifiers, either for voxelwise airway classification~\cite{Lo2010,Bian2018} or to remove false positive airway candidates from a leaky baseline segmentation~\cite{Inoue2013,Meng2016}. These classifiers (kNN~\cite{Lo2010}, AdaBoost~\cite{Inoue2013}, support vector machines~\cite{Meng2016} or random forest~\cite{Bian2018}) operate on a set of predefined image features: multiscale Gaussian derivatives~\cite{Lo2010}, multiscale Hessian-based features~\cite{Inoue2013,Meng2016,Bian2018} or image texture features with local binary patterns~\cite{Bian2018}. These methods can achieve more complete airway tree predictions than previous purely intensity-based methods, with fewer false positives. However, they are highly dependent on the image features used to train the classifier, and may be time-consuming to apply due to the time needed to compute the image features~\cite{Lo2010,Meng2016,Bian2018}. Recently, many state-of-the-art methods for medical image segmentation tasks have used deep learning~\cite{Litjens2017}, and in particular convolutional neural networks (CNNs)~\cite{Long2015}. The main advantage of deep CNN methods over classical learning-based techniques is that the extraction of relevant image features is done automatically from data in an end-to-end optimised setting. Several CNN-based methods have been applied for airway segmentation~\cite{Charbonnier2017,Yun2019,Jin2017,Meng2017,GarciaUceda2018,Qin2019,Qin2019-2,Zhao2019,Wang2019,Qin2020,Qin2021,Zheng2020,Zhou2021,Nadeem2021,GarciaUceda2019}. Charbonnier et al.~\cite{Charbonnier2017} applied CNNs to detect and remove leakage voxels from a leaked region growing based segmentation. Yun et al.~\cite{Yun2019} applied the so-called 2.5D CNN, a pseudo 3D CNN which processes the three perpendicular 2D slices around each voxel, to perform voxelwise airway classification. The methods~\cite{Jin2017,Meng2017,GarciaUceda2018,GarciaUceda2019,Qin2019,Qin2019-2,Zhao2019,Wang2019,Qin2020,Qin2021,Zheng2020,Zhou2021,Nadeem2021} are based on the U-Net architecture~\cite{Ronneberger2015}. This model and its 3D extension~\cite{Cycek2016,Milletari2016} are highly successful for medical image segmentation. The main advantage of the U-Net over voxelwise CNN models is that it can process entire images in one forward pass through its encoding / decoding structure, resulting in a segmentation map directly. Jin et al.~\cite{Jin2017} and Meng et al.~\cite{Meng2017} applied the U-Net on local volumes of interest around airways, being guided by the centerlines from a baseline segmentation~\cite{Jin2017} or by tracking the airways extracted from the U-Net~\cite{Meng2017}. Garcia-Uceda et al.~\cite{GarciaUceda2018} applied the U-Net on large image patches extracted from the 3D scans, using various data augmentation techniques. Garcia-Uceda et al.~\cite{GarciaUceda2019} proposed a joint approach combining both 3D U-Net and graph neural networks (GNNs)~\cite{Selvan2020}. Qin et al.~\cite{Qin2019,Qin2019-2} applied the U-Net for 26-fold prediction of the 26-neighbour connectivities of airway voxels, and segments airways by grouping voxels with at least one mutually predicted connectivity with a neighbour voxel. Zhao et al.~\cite{Zhao2019} combined both 2D and 3D U-Nets with linear programming-based tracking to link disconnected components of the segmented airway tree. Wang et al.~\cite{Wang2019} used the U-Net with spatial recurrent convolutional layers and a radial distance loss to better segment tubular structures. Qin et al.~\cite{Qin2020,Qin2021} extended the U-Net with feature recalibration and attention distillation modules that leverage the knowledge from intermediate feature maps of the network. Zheng et al.~\cite{Zheng2020} proposed a General Union loss to alleviate the intra-class imbalance between large and small airways. Zhou et al.~\cite{Zhou2021} extended the U-Net with a multi-scale feature aggregation module based on dilated convolutions, to include more contextual information from a larger receptive field. Nadeem et al.~\cite{Nadeem2021} used the U-Net followed by a freeze-and-grow propagation algorithm to iteratively increase the completeness of the segmented airway tree.

In this paper, we present a fully automatic and end-to-end optimised method to perform airway segmentation from chest CTs, based on the U-Net architecture. A preliminary version of this work was presented in~\cite{GarciaUceda2018}. The proposed method processes larger 3D image patches, often covering an entire lung, in a single pass through the network. We achieve this by using a simple U-Net backbone with low memory footprint, and having efficient operations that feed image data to the network. This makes our method simple, robust and efficient. We performed a thorough validation of the proposed method on three datasets with very different characteristics, from subjects including pediatric patients and adults and with diverse airway abnormalities, including the EXACT'09 public dataset~\cite{LoExact2012}. We compared the method with other state-of-the-art segmentation methods tested on the same data. On the EXACT'09 data, we compared with all the recent methods in the literature evaluated on these data, in terms of the average performance measures reported by the EXACT'09 evaluation. On the other datasets, we compared with the two methods previously evaluated on the same data~\cite{Lo2010,Lo2009}. Moreover, we evaluated the accuracy of our method with respect to the presence of lung disease and airway abnormalities in the CT scans from the three datasets.

This article is organized as follows: in Section 2 we explain the proposed methodology for airway segmentation, in Section 3 we present the data used for the experiments in this paper, in Section 4 we explain the experiments performed to validate the method, in Section 5 we present the results obtained, in Section 6 we discuss these results and explain the advantages and limitations of the proposed method, and in Section 7 we give the conclusions for this paper.


\section{Methods}
\label{sec:Methods}

\subsection{Network architecture}
\label{sec:NetsArchitecture}

\begin{figure}[t]
\centering
\begin{subfigure}[t]{\textwidth}
\includegraphics[width=\textwidth]{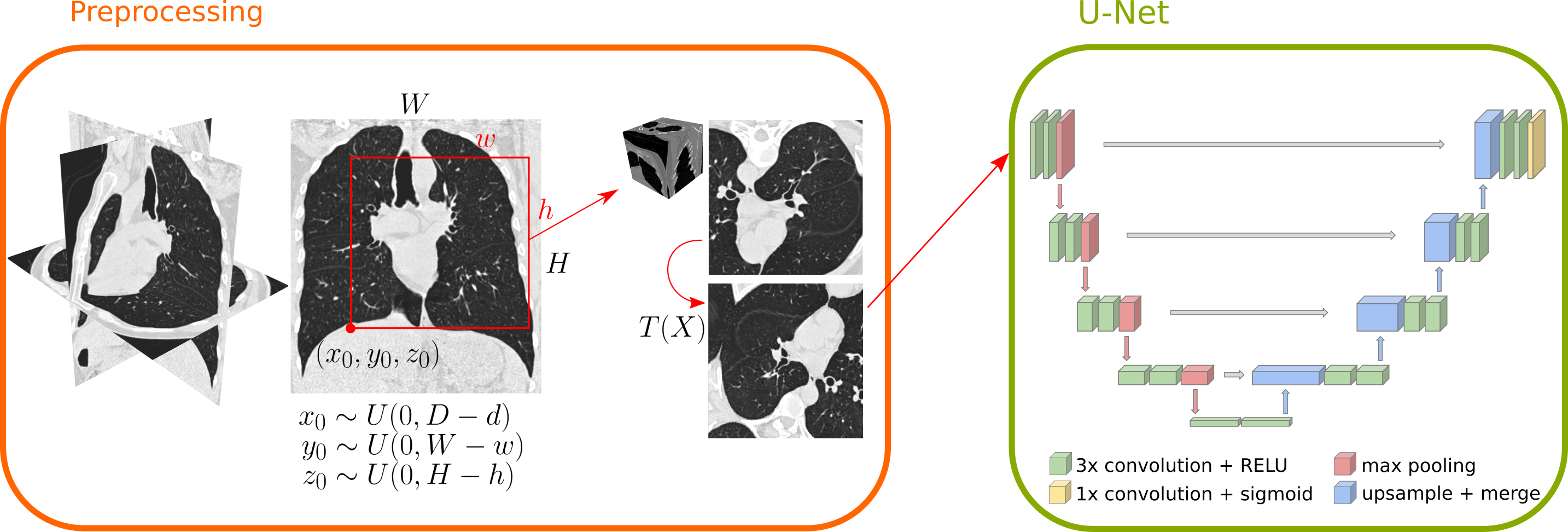}
\caption{Training time}
\end{subfigure}
\bigbreak
\centering
\begin{subfigure}[t]{\textwidth}
\includegraphics[width=\textwidth]{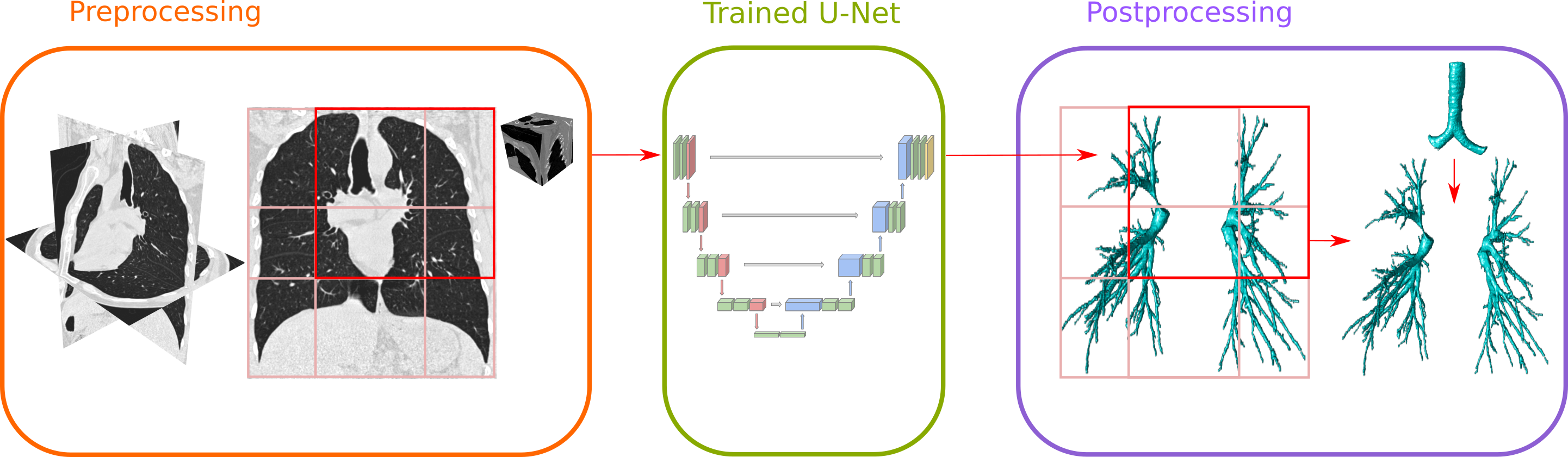}
\caption{Testing time}
\end{subfigure}
\caption{Schematics of the proposed airway segmentation method. (a) Overview at training time, showing the extraction of 3D random patches from the CT scan, to feed as input to the U-Net. (b) Overview at testing time, showing the extraction of 3D patches through sliding-window, and the generation of the airway tree segmentation from the patch-wise output of the U-Net.}
\label{figAirwayMethod}
\end{figure}

The network consists of an encoder (downsampling) path followed by a decoder (upsampling) path, at different levels of resolution. Each level in the downsampling / upsampling path is composed of two $3\timestxt3\timestxt3$ convolutional layers, each followed by a rectified linear (ReLU) activation, and a $2\timestxt2\timestxt2$ pooling / upsample layer, respectively. After each pooling / upsample layer, the number of feature channels is doubled / halved, respectively. There are skip connections between opposing convolutional layers prior to pooling and after upsampling, at the same resolution level. The final layer is a $1\timestxt1\timestxt1$ convolutional layer followed by a sigmoid activation function. This network is schematically shown in Fig.~\ref{figAirwayMethod}.

For analysis of 3D chest scans, we found that the main bottleneck is the memory footprint of the network. To alleviate this, we use non-padded convolutions in the first 3 resolution levels of the U-Net, where the outmost layers of voxels in the feature maps are progressively removed after each convolution. We still use zero-padding in the remaining levels after the third, to prevent an excessive reduction of the output size of the network. This allows to reduce the memory footprint by approximately $30\%$ compared to a model with zero-padding in all layers. Moreover, we do not use dropout or batch normalization layers, as these require extra memory to store the feature maps after the operation.

\subsection{Training of networks}
\label{sec:TrainingNets}
The network is trained in an end-to-end supervised fashion, using the CT scans as input and evaluating the network output with respect to the ground truth, the reference airway segmentations, using the soft Dice loss~\cite{SoftDice2017},
\begin{equation}
\centering
\mathcal{L} = 1 - \frac{2 \sum_{x\in N_L}{p(x) g(x)}}{\sum_{x\in N_L}{p(x)} + \sum_{x\in N_L}{g(x)}}
\label{eq:trainloss}
\end{equation}
where $p(x), g(x)$ are the predicted voxelwise airway probability maps and airway ground truth, respectively. In the loss computation, the ground truth is masked to the region of interest (ROI): the lungs, indicated by the sub-index $L$, so that only voxels within this region contribute to the loss. This mask removes the trachea and part of the main bronchi from the ground truth, so that the training of the network focuses on the smaller branches. The lung segmentation needed for this masking operation is easily computed by a region growing method~\cite{Lo2010}.

\subsection{Implementation of networks}
\label{sec:NetsImplementation}

When building the U-Net, the degrees of freedom (or hyperparameters) that determine the capacity of the network are i) the number of resolution levels in the U-Net, ii) the number of feature maps in the first layer, iii) the input image size, and iv) the training batch size. We optimized these hyperparameters using subsets of training data from the two datasets used in this paper. We observed that the input image size was the most important parameter, and that by using larger input patches we obtained better results and faster convergence of the training and validation losses. With this, we found an optimal U-Net of 5 levels, 16 feature maps in the first layer, input size of $252\timestxt252\timestxt252$, and training batches containing only one image. This network can fit in a graphical card GPU NVIDIA Quadro P6000 with 24 GB memory during training. This U-Net construction is used for all the experiments undertaken in this paper. The implementation of the network is done using the Pytorch framework~\cite{Pytorch2019}.

\subsection{Generation of images during training}
\label{sec:PreprocessImages}
The 3D chest scans analysed in this paper have a size much larger than the input size of the network. Thus, we extract random patches of size $252\timestxt252\timestxt252$ from the full-size scans and ground truth to train the network. Moreover, the training datasets used for the experiments in this paper are small and contain only 15-25 scans. Thus, we apply data augmentation to the extracted image patches to increase artificially the amount of training data available and improve generalization. To generate input images to the network from the chest scans and ground truth, we apply the series of operations:

\begin{enumerate}[itemsep=0pt]
\item Crop the full-size scans to a bounding-box around the region of interest: the pre-segmented lung fields. This operation removes the areas outside the lungs, which are irrelevant for airway segmentation. The limits of the bounding-box are defined as 30 voxels from the outer voxel of the lung mask, in each direction. The extra 30-voxel buffer is used to prevent that boundary effects introduced by the use of zero-padding in the last layers of the network affect the prediction of peripheral airways, closer to the lung surface. Moreover, we mask the ground truth to the mask of the lung fields to remove the trachea and part of the main bronchi.
\item Extract image patches from the input image by cropping this to random bounding-boxes of $252\timestxt252\timestxt252$. We generate a total of 8 random patches per scan and per epoch during training. To perform this operation, we randomly sample the coordinates of the first corner of the bounding-box $(x_0, y_0, z_0)$ from a uniform distribution in the ranges $x_0\in[0, D-d]$, $y_0\in[0, W-w]$ and $z_0\in[0, H-h]$, respectively, where $(d, w, h)$ is the size of the extracted patch, and $(D, W, H)$ the size of the cropped input image. Then, the random bounding-box is generated with the limits $((x_0, x_0+d), (y_0, y_0+w), (z_0, z_0+h))$.
\item Apply random rigid transformations to the input image patches as data augmentation. These transformations include i) random flipping in the three directions, ii) random small 3D rotations up to 10 degrees, and iii) random scaling with a scale factor in the range (0.75, 1.25). The same transformations are applied to the ground truth using nearest-neighbour interpolation. 
\end{enumerate}

This series of operations is schematically shown in Fig.~\ref{figAirwayMethod}. We implemented operations 2) and 3) as an efficient on-the-fly image generation pipeline that feed data to the network, to reduce any computational overhead for the batch data loading part of the training algorithm. Operation 1) is applied only once to the chest scans and ground truth and prior to the training.

\subsection{Airway extraction}
\label{sec:PostprocessImages}
To segment new scans, we extract image patches of size $252\timestxt252\timestxt252$ from the full-size scans in a sliding-window fashion in the three dimensions, with an overlap of $50\%$ between the patches in each direction. Each image patch is processed by the trained network, producing a set of patch-wise airway probability maps. To obtain the full-size binary segmentation of the airway tree from these output patches, we apply the series of operations:

\begin{enumerate}[itemsep=0pt]
\item Aggregate the output patches by the network with airway probability maps by stitching them together (i.e. reversing the sliding-window operation used to extract the input patches). To account for the overlap between patches, we divide the aggregated voxelwise airway probabilities by the number of patches containing the given voxel.
\item Mask the full-size airway probability map to the mask of the lung fields. This is to discard the noise predictions outside the lung regions, as only these regions are included in the training of the networks by equation~\eqref{eq:trainloss}.
\item Threshold the airway probability maps to obtain the airway binary segmentation, using a threshold of 0.5 by default. This output does not contain the trachea and part of the main bronchi that were outside the mask of the lung fields. To include this, we merge the airway segmentation with a mask for the trachea and main bronchi. This is easily computed by a region growing method~\cite{Lo2010}, and then masking its output to the mask of the lung fields.
\item Apply connected component analysis~\cite{ConnCompon1996} to the airway segmentation and retrieve the final airway tree as the largest 26-neighbour-connected component.
\end{enumerate}


\section{Data}
\label{sec:Data}

To conduct the experiments, we used CT scans and reference airway segmentations from three different datasets:
\begin{enumerate}

\item \textbf{CF-CT}: This dataset consists of 24 inspiratory low-dose chest CT scans from pediatric patients at Erasmus MC-Sophia, 12 controls and 12 diseased~\cite{Kuo2017}. The 12 controls were patients with normal lung assessment from CT reported by two different radiologists, who received CT scanning for diagnostic purposes. The 12 diseased were 11 patients with Cystic fibrosis (CF) lung disease and 1 with Common Variable Immune Deficiency (CVID), who had structural lung damage with airway abnormalities visible on the CT scan. The two groups were gender and age matched, with the age range from 6 to 17 years old in both groups and 5 females per group. Scanning was undertaken using spirometry-guidance in a Siemens SOMATOM Definition Flash scanner. Each CT scan has an in-plane voxel size in the range $0.44-0.71$ mm, with slice thickness in the range $0.75-1.0$ mm, and slice spacing from 0.3 to 1.0 mm.

\item \textbf{DLCST}: This dataset consists of 32 inspiratory low-dose chest CT scans from the Danish Lung Cancer Screening Trial~\cite{DLCST2009}. Participants to the trial were subjects between 50 to 70 years old, with a history of at least 20 pack-years of smoking, and without lung cancer related symptoms. Roughly half of the CT scans are from subjects with Chronic Obstructive Pulmonary Disease (COPD), reported from Spirometry measures. Scanning was undertaken using a MDCT scanner (16 rows Philips Mx 8000, Philips Medical Systems, Eindhoven, The Netherlands). All CT scans have a voxel resolution of $0.78\timestxt0.78\timestxt1$ mm$^3$.

\item \textbf{EXACT'09}: This is the multi-center, public dataset of the EXACT'09 airway extraction challenge~\cite{LoExact2012}. We used for evaluation purposes the EXACT'09 test set consisting of 20 chest CT scans. The conditions of the scanned patients vary largely, ranging from healthy volunteers to subjects with severe lung disease. The data includes both inspiratory and expiratory scans, ranging from clinical to low-dose. The CT scans were acquired with several different CT scanner brands and models, using a variety of scanning protocols and reconstruction parameters. Each CT scan has an in-plane voxel resolution in the range $0.55-0.78$ mm, with varying slice thickness in the range $0.5-1.0$ mm.
\end{enumerate}

\subsection{Generation of reference segmentations}
\label{sec:GenerationGroundtruth}

Networks were trained and evaluated using airway lumen segmentations that were obtained by a combination of manual interaction and automated surface detection. For the CF-CT data, we have centerline annotations manually drawn by trained experts. For the DLCST data, we have airway extractions obtained from the union of the results of methods~\cite{Lo2010} and~\cite{Lo2009}, and corrected by a trained observer. The visual assessment was done similarly to the EXACT'09 challenge~\cite{LoExact2012}, where the obtained airways trees were divided in branches using a wave front propagation algorithm that detects bifurcations, and spurious branches were removed manually. "Partly wrong" branches whose segmentation covered the airway lumen but leaked outside the airway wall were also removed. To obtain the ground truth airway segmentations, we applied the optimal surface graph segmentation method~\cite{Petersen2014} on top of these initial airway references in order to i) determine accurate lumen contours and ii) homogenize the ground truth for the two datasets. To evaluate the networks, we use these two ground truth segmentations as well as (via submission to the EXACT'09 challenge) the reference segmentations from the EXACT'09 test set. To build the EXACT'09 reference, the airway trees predicted by all 15 participating methods to the challenge were evaluated by independent expert observers, and the group of correct-scored branches were merged together.


\section{Experiments}
\label{sec:Experiments}

We performed three experiments to assess the performance of the proposed U-Net-based method in the different datasets and in comparison with previously published approaches:
\begin{enumerate}

\item \textbf{CF-CT}: We compared the performance of our method with that of a previously published airway segmentation method on these data~\cite{PerezRovira2016}. These results were obtained by a kNN-classifier together with a vessel similarity measure~\cite{Lo2010}, and refined with an optimal surface graph method~\cite{Petersen2014} to obtain an accurate lumen segmentation. We denote this method by~\textbf{kNN-VS}. We trained and evaluated our U-Net model using the same 6-fold cross-validation setting as in~\cite{PerezRovira2016}. The training and testing data in each fold have an equibalanced number of scans from control and diseased subjects.

\item \textbf{DLCST}: We compared the performance of our method with that of two previously published airway segmentation methods on these data~\cite{Lo2010} and~\cite{Lo2009}. The method~\cite{Lo2010} is a kNN-classifier together with a vessel similarity measure, while the method~\cite{Lo2009} extends the airways iteratively from an incomplete segmentation with locally optimal paths. We refined these results~\cite{Lo2010,Lo2009} with an optimal surface graph method~\cite{Petersen2014} to obtain a more accurate lumen segmentation, and to homogenize them with the training data used to train our model, for a fair comparison. We denote these methods by~\textbf{kNN-VS} and~\textbf{LOP}, respectively. We trained and evaluated our U-Net model using the same 2-fold cross-validation setting as in~\cite{Lo2010} and~\cite{Lo2009}. The training and testing data in each fold have a similar number of scans from healthy and diseased subjects.

\item \textbf{EXACT'09}: We compared the performance of our method with that of the 15 methods participating in the challenge EXACT'09 based on the results reported in~\cite{LoExact2012}, 6 post-challenge methods evaluated on the EXACT'09 data and reported in~\cite{EXACTwebsite}, and 4 recent airway segmentation methods evaluated on these data~\cite{Charbonnier2017,Yun2019,Gil2019,Qin2021}. The methods in Charbonnier et al.~\cite{Charbonnier2017}, Yun et al.~\cite{Yun2019} and Qin et al.~\cite{Qin2021} were previously described in the Introduction. The method in Gil et al.~\cite{Gil2019}, named PICASSO, uses locally adaptive optimal thresholds learnt from a graph-encoded measure of the airway tree branching. We denote these methods by~\textbf{CNN-Leak},~\textbf{2.5D CNN},~\textbf{U-Net-FRAD} and ~\textbf{PICASSO}, respectively. The method in Zheng et al.~\cite{Zheng2020} is also evaluated on EXACT'09 data, but we did not include it in our comparison as the authors did not report the same specificity metric (false positive rate) as in the EXACT'09 evaluation. To our knowledge, these are all the recent methods in the literature that are evaluated on the EXACT'09 data. We trained our U-Net model using half of the CT scans from the CF-CT and DLCST data (28 scans in total). This training data has a similar number of scans from healthy and diseased subjects. We did not include the EXACT'09 training set in our training data as there are no reference segmentations available. Additionally, we compared our method with the nnU-Net segmentation method proposed by Isensee et al.~\cite{nnUnet2021}, applied to airway segmentation. We denote this method by~\textbf{nnU-Net}. To do this, we performed an experiment with the nnU-Net model: we trained it using the same training data as for our method, and we evaluated it on EXACT'09 data following the same evaluation protocol. We provide implementation details of the nnU-Net method and of our experiment in the Appendix.

\end{enumerate}

For each cross-validation fold in the experiments, we used roughly 80\% of images in the training fold to train the models, and the remaining 20\% is used to monitor the validation loss. To train the models, we used the Adam optimizer~\cite{AdamOptime2017} with an initial learning rate of $10^{-4}$, which was chosen as large as possible while ensuring convergence of the training and validation losses. As convergence criterion, we monitored the moving average of the validation loss over 50 epochs, and training is stopped when its value i) increases by more than 5\%, or ii) does not decrease more than 0.1\%, over 20 epochs. We trained the models until convergence is reached, and we retrieved the model with the overall minimum validation loss for testing. Training time was approximately 1-2 days on a GPU NVIDIA Quadro P6000, while test time inference takes less than 1 min per scan including all post-processing steps to obtain the binary airway tree.

To compute the airway predictions on the EXACT'09 data, we thresholded the airway probability maps with a value of 0.1, and retrieved the final airway tree as the largest 6-neighbour-connected component. The lower threshold was to compensate for the reduction of completeness in our results when computing a single 6-connected structure, required by the submission guidelines for the evaluation on the EXACT'09 dataset~\cite{LoExact2012}. In contrast, for the predictions on the CF-CT and DLCST data, we used the default threshold of 0.5 and 26-neighbour-connected component analysis.

\subsection{Evaluation}
\label{sec:EvaluationMetrics}

To compare the different methods on the CF-CT and DLCST datasets, we computed i) tree length detected, ii) centerline leakage and iii) Dice coefficient, with respect to the ground truth segmentations. For the results on the EXACT'09 data, we evaluated i) tree length detected and ii) false positive rate, computed by the EXACT'09 challenge organizers~\cite{LoExact2012}. Moreover, we assessed the segmentation accuracy of our U-Net-based method with respect to the presence of lung disease in the scans from each dataset. For this, we computed i) tree length detected, ii) centerline leakage (or false positive rate on the EXACT'09 data) and iii) total tree length, and compared the measures from both healthy and diseased subjects.

The metrics are defined as:
\begin{enumerate}[itemsep=0pt]
\item \textbf{Tree length (TL)}: The number of ground truth centerline voxels that are inside the predictions, as a percentage of the ground truth centerline full length.
\item \textbf{Centerline leakage (CL)}: The number of predicted centerline voxels outside the ground truth, as a percentage of the ground truth centerline full length.
\item \textbf{False positive rate (FPR)}: The number of false positive voxels outside the ground truth, as a percentage of the total number of ground truth voxels.
\item \textbf{Dice coefficient (DSC)}: The voxelwise overlap between the predictions and the ground truth according to equation~\eqref{eq:dicecoeff}, with $P$ and $G$ the binary prediction and ground truth masks, respectively.
\begin{equation}
\centering
DSC = \frac{2 |P \cap G|}{|P| + |G|}
\label{eq:dicecoeff}
\end{equation}
\item \textbf{Total tree length}: The sum of ground truth centerline voxels that are inside the predictions, multiplied by a factor that represents the voxel resolution (we used the geometrical mean of the voxel sizes in the three dimensions).
\end{enumerate}

The trachea and main bronchi are removed in these measures from both the predictions and ground truth, similar to~\cite{LoExact2012}. The centerlines are obtained by applying skeletonization~\cite{Skeletonize1994} to the prediction and ground truth masks. 

To compare the performance of the different methods, we used the paired two-sided Student's T-test on the performance measures over the test set. To compare the accuracy of our U-Net-based method between healthy and diseased subjects, we used the unpaired two-sided Student's T-test on the measures from the two groups. We consider that a p-value lower than 0.01 indicates that the two sets of measures compared are significantly different.


\section{Results}
\label{sec:Results}

\subsection{Evaluation on CF-CT data}
\label{sec:ResultsCFCT}

\begin{figure}[t]
\centering
\begin{subfigure}[t]{0.33\textwidth}
\includegraphics[width=\textwidth]{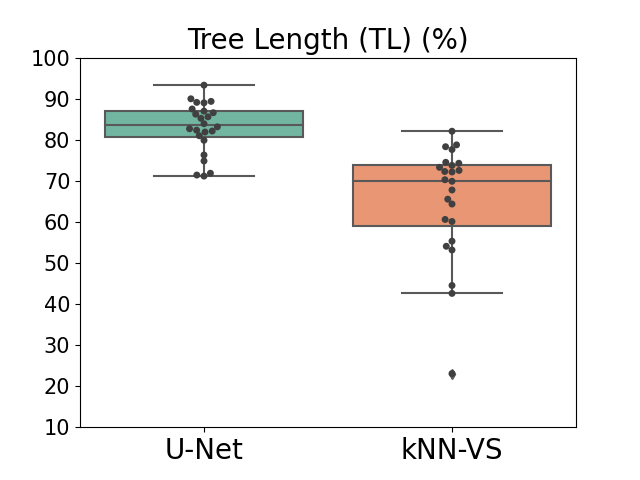}
\end{subfigure}
\begin{subfigure}[t]{0.33\textwidth}
\includegraphics[width=\textwidth]{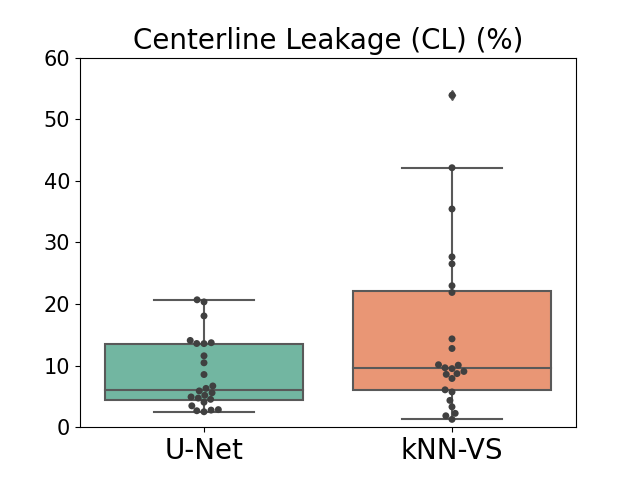}
\end{subfigure}
\begin{subfigure}[t]{0.33\textwidth}
\includegraphics[width=\textwidth]{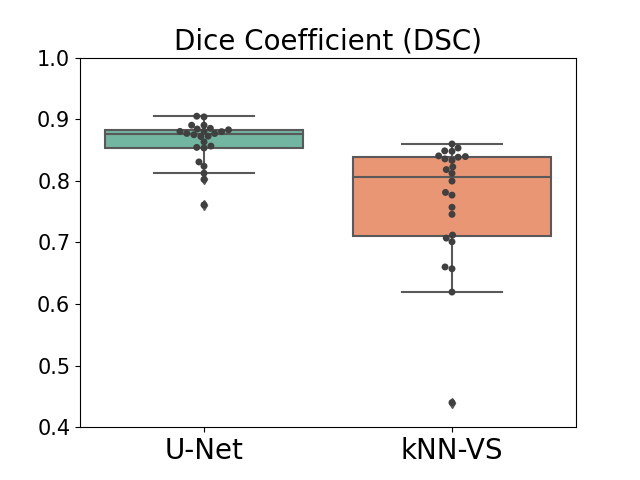}
\end{subfigure}
\caption{Boxplots showing i) tree length detected, ii) centerline leakage and iii) Dice coefficient on the CF-CT data, for the results obtained by our U-Net-based method and the kNN-VS method~\cite{PerezRovira2016}. For each boxplot, the box shows the quartiles of the data (defined by the median, 25\% and 75\% percentiles), the whiskers extend to include the data within 1.5 times the interquartile range from the box limits, and the markers show all the data points.}
\label{figBoxplotResCFCT}
\end{figure}

\begin{figure}[t]
\centering
\includegraphics[width=\textwidth]{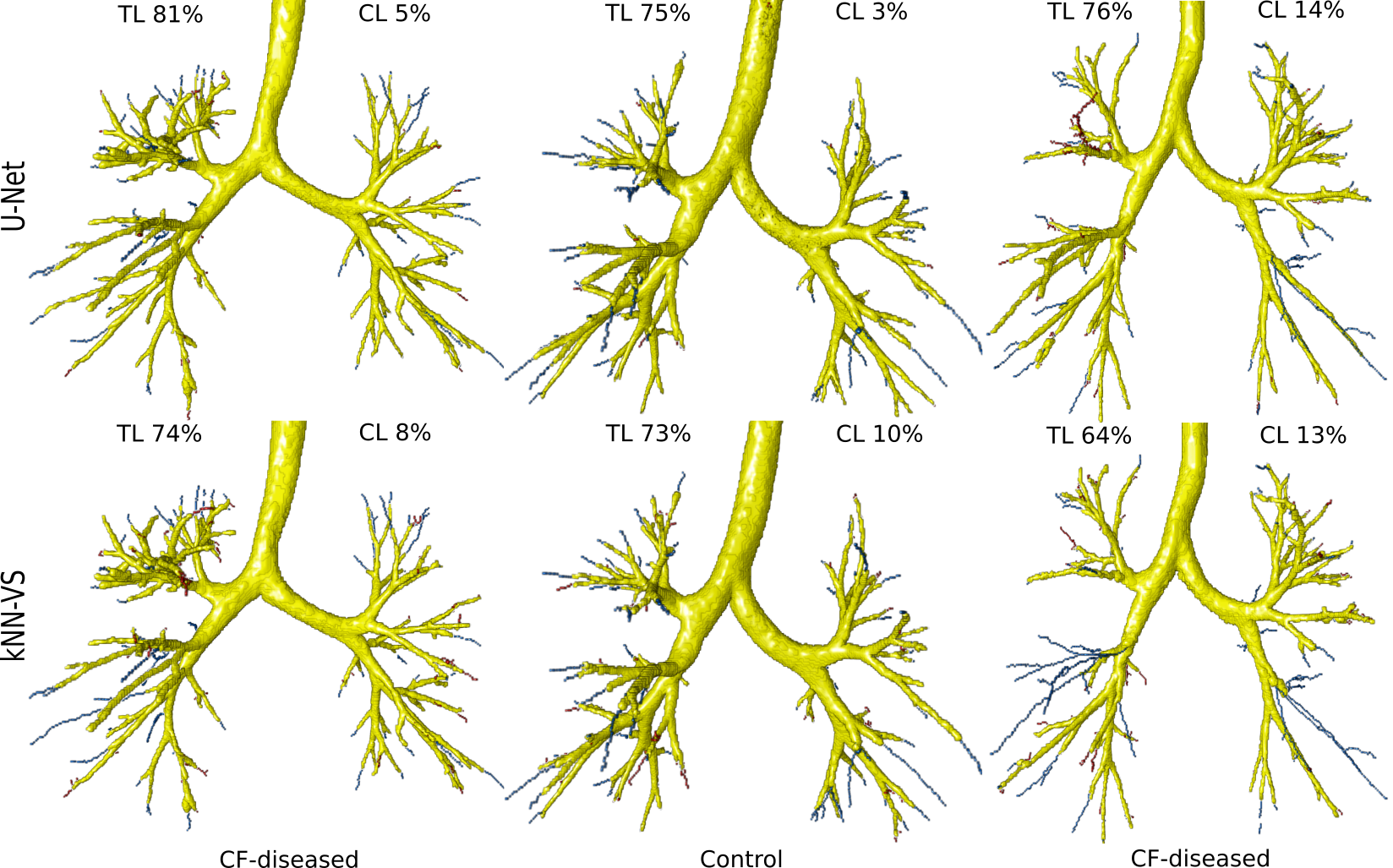}
\caption{Visualization of predicted airways trees on the CF-CT data, obtained by our U-Net-based method and the kNN-VS method~\cite{PerezRovira2016}. True positives are displayed in yellow, centerline false negatives in blue and centerline false positives in red. These false negatives and positives are the errors counted in the tree length detected and centerline leakage measures reported in Fig.~\ref{figBoxplotResCFCT}, respectively. The cases displayed correspond to, from left to right, the two U-Net results with Dice score closest to the median of the test Dice measures, and the U-Net result with the lowest Dice score.}
\label{figVisualResCFCT}
\end{figure}

We show in Fig.~\ref{figBoxplotResCFCT} the evaluation on the CF-CT data of our U-Net-based method and the kNN-VS method~\cite{PerezRovira2016}. Comparing the U-Net with the kNN-VS results, the former shows higher TL ($83.5(80.7-87.1)$ compared to $70.1(58.9-73.9)$, $p<0.001$), higher DSC ($0.876(0.854-0.883)$ compared to $0.806(0.711-0.839)$, $p<0.001$) and a similar CL ($6.09(4.41-13.6)$ compared to $9.58(5.98-22.1)$, $p=0.021$). This indicates that our U-Net-based method predicts more complete airway trees than the kNN-VS method~\cite{PerezRovira2016}, with more and/or longer peripheral branches, and with a similar count of false positive errors. We show in Fig.~\ref{figVisualResCFCT} visualizations of airway trees obtained by the two methods. We can see that the airways predicted by our U-Net-based method match better the reference segmentations, with more and/or longer peripheral branches detected, while the kNN-VS method misses many of these peripheral branches.

Comparing the measures from our U-Net-based method between control and subjects with CF and airway abnormalities, we found no significant differences in TL ($84.1(78.6-87.9)$ compared to $83.5(81.7-86.4)$, $p=0.87$) and in CL ($5.74(3.91-13.8)$ compared to $6.45(4.68-11.2)$, $p=0.78$), but they were significant in total tree length ($146(135-156)$ compared to $263(208-308)$, $p<0.01$). This indicates that our method has similar accuracy with respect to the manual annotations on scans from both control and subjects with CF, while segmenting more airway branches in the diseased cases. We show these grouped measures in the Appendix.

\subsection{Evaluation on DLCST data}
\label{sec:ResultsDLCST}

\begin{figure}[t]
\centering
\begin{subfigure}[t]{0.33\textwidth}
\includegraphics[width=\textwidth]{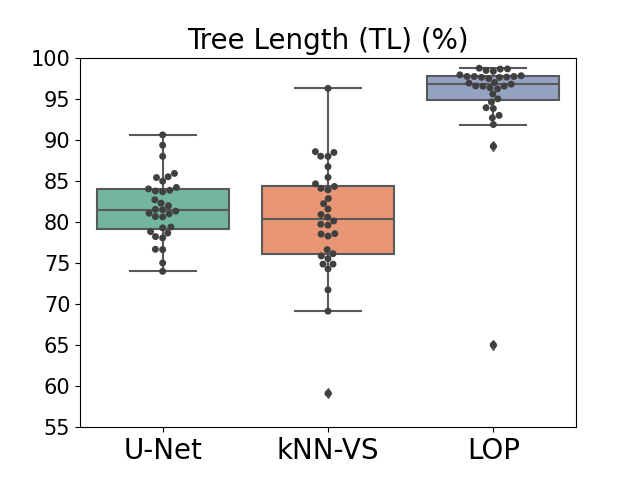}
\end{subfigure}
\begin{subfigure}[t]{0.33\textwidth}
\includegraphics[width=\textwidth]{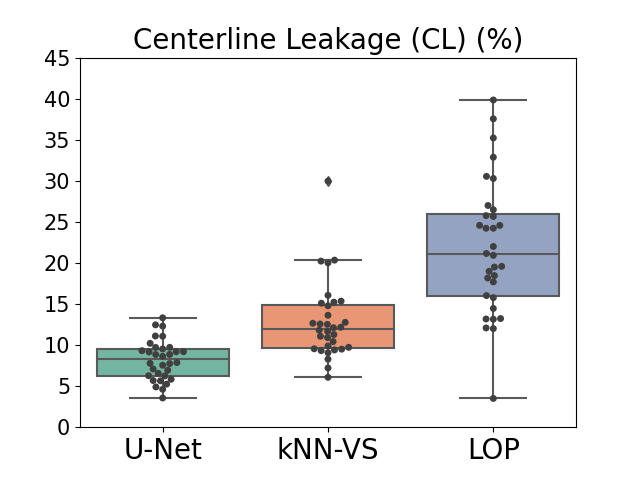}
\end{subfigure}
\begin{subfigure}[t]{0.33\textwidth}
\includegraphics[width=\textwidth]{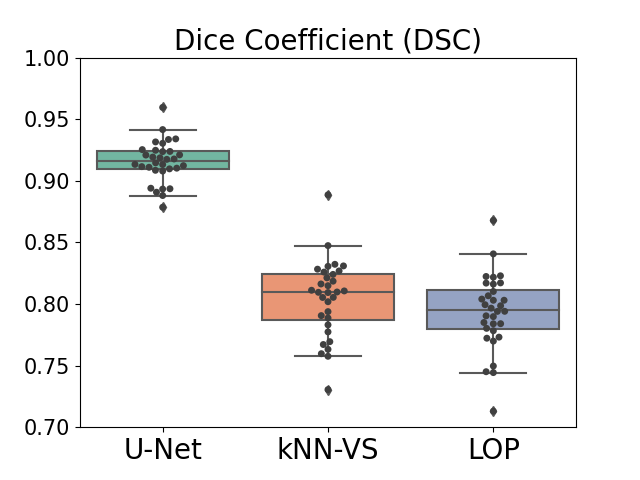}
\end{subfigure}
\caption{Boxplots showing i) tree length detected, ii) centerline leakage and iii) Dice coefficient on the DLCST data, for the results obtained by our U-Net-based method and the kNN-VS~\cite{Lo2010} and LOP~\cite{Lo2009} methods. For each boxplot, the box shows the quartiles of the data (defined by the median, 25\% and 75\% percentiles), the whiskers extend to include the data within 1.5 times the interquartile range from the box limits, and the markers show all the data points.}
\label{figBoxplotResDLCST}
\end{figure}

\begin{figure}[t!]
\centering
\includegraphics[width=\textwidth]{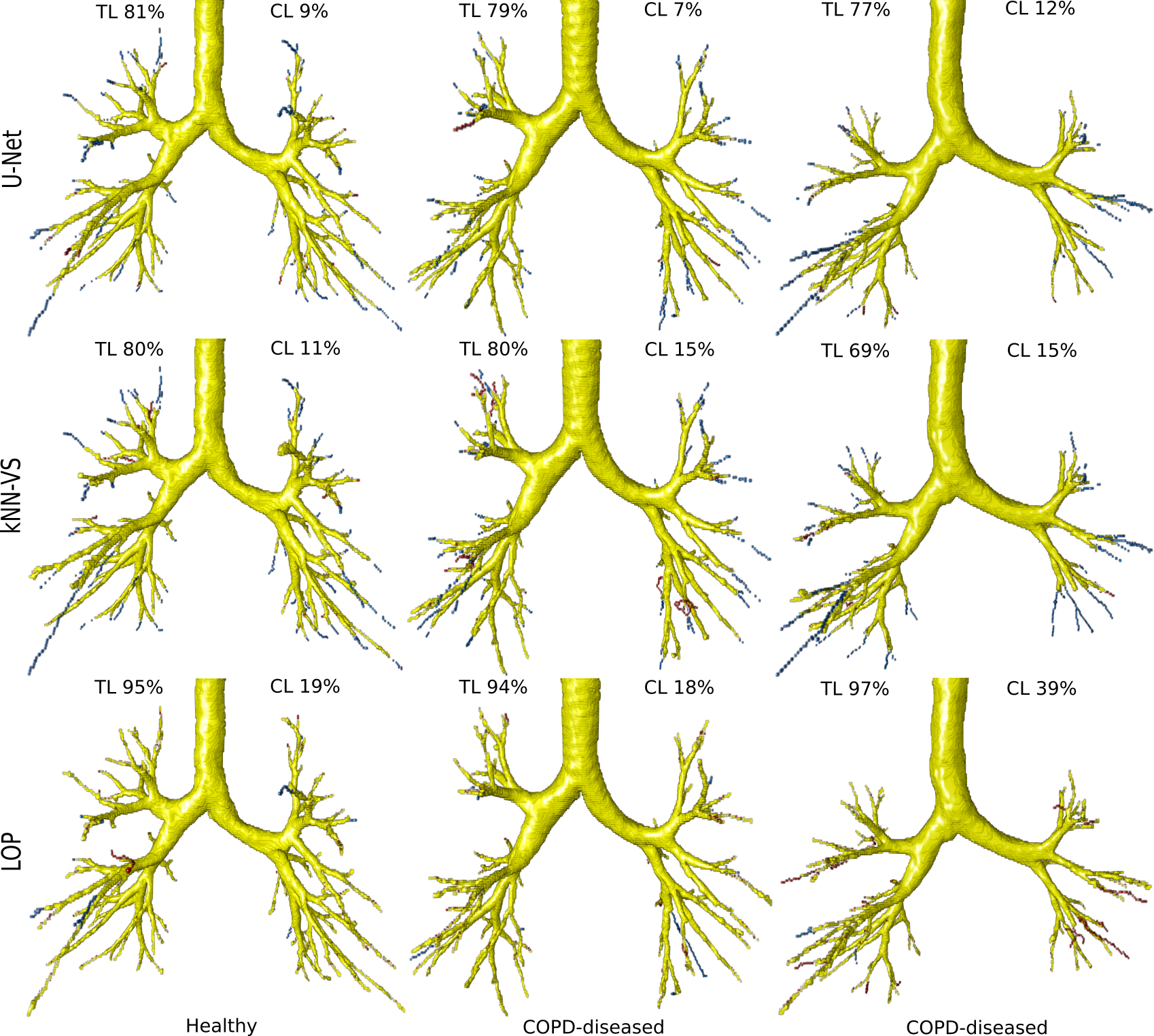}
\caption{Visualization of predicted airways trees on the DLCST data, obtained by our U-Net-based method and the kNN-VS~\cite{Lo2010} and LOP~\cite{Lo2009} methods. True positives are displayed in yellow, centerline false negatives in blue and centerline false positives in red. These false negatives and positives are the errors counted in the tree length detected and centerline leakage measures reported in Fig.~\ref{figBoxplotResDLCST}, respectively. The cases displayed correspond to, from left to right, the two U-Net results with Dice score closest to the median of the test Dice measures, and the U-Net result with the lowest Dice score.}
\label{figVisualResDLCST}
\end{figure}

We show in Fig.~\ref{figBoxplotResDLCST} the evaluation on the DLCST data of our U-Net-based method and the kNN-VS~\cite{Lo2010} and LOP~\cite{Lo2009} methods. Comparing the U-Net with the kNN-VS results, the former shows lower CL ($8.25(6.26-9.58)$ compared to $12.0(9.68-14.9)$, $p<0.001$), higher DSC ($0.916(0.909-0.924)$ compared to $0.810(0.787-0.825)$, $p<0.001$) and a similar TL ($81.5(79.2-84.1)$ compared to $80.3(76.1-84.4)$, $p=0.19$). Comparing the U-Net with the LOP results, the former shows lower CL ($8.25(6.26-9.58)$ compared to $21.0(16.0-25.9)$, $p<0.001$), higher DSC ($0.916(0.909-0.924)$ compared to $0.795(0.780-0.812)$, $p<0.001$), but lower TL ($81.5(79.2-84.1)$ compared to $96.8(94.9-97.7)$, $p<0.001$). This indicates that our U-Net-based method predicts an airway tree as complete as the kNN-VS method~\cite{Lo2010}, and less complete than the LOP method~\cite{Lo2009}, but with significantly less false positive errors than these two methods. It should be noted that this comparison is biased towards the kNN-VS and LOP methods that made up the reference segmentations, which partly explains the very high completeness of the results by the LOP method. We show in Fig.~\ref{figVisualResDLCST} visualizations of airway trees obtained by the three methods. We can see that the airways predicted by the LOP method have more peripheral branches detected (with almost no false negative errors), however they have more false positive errors. The results from our U-Net-based method have less peripheral branches detected, but also have much less false positives.

\begin{figure}[t]
\centering
\includegraphics[width=0.8\textwidth]{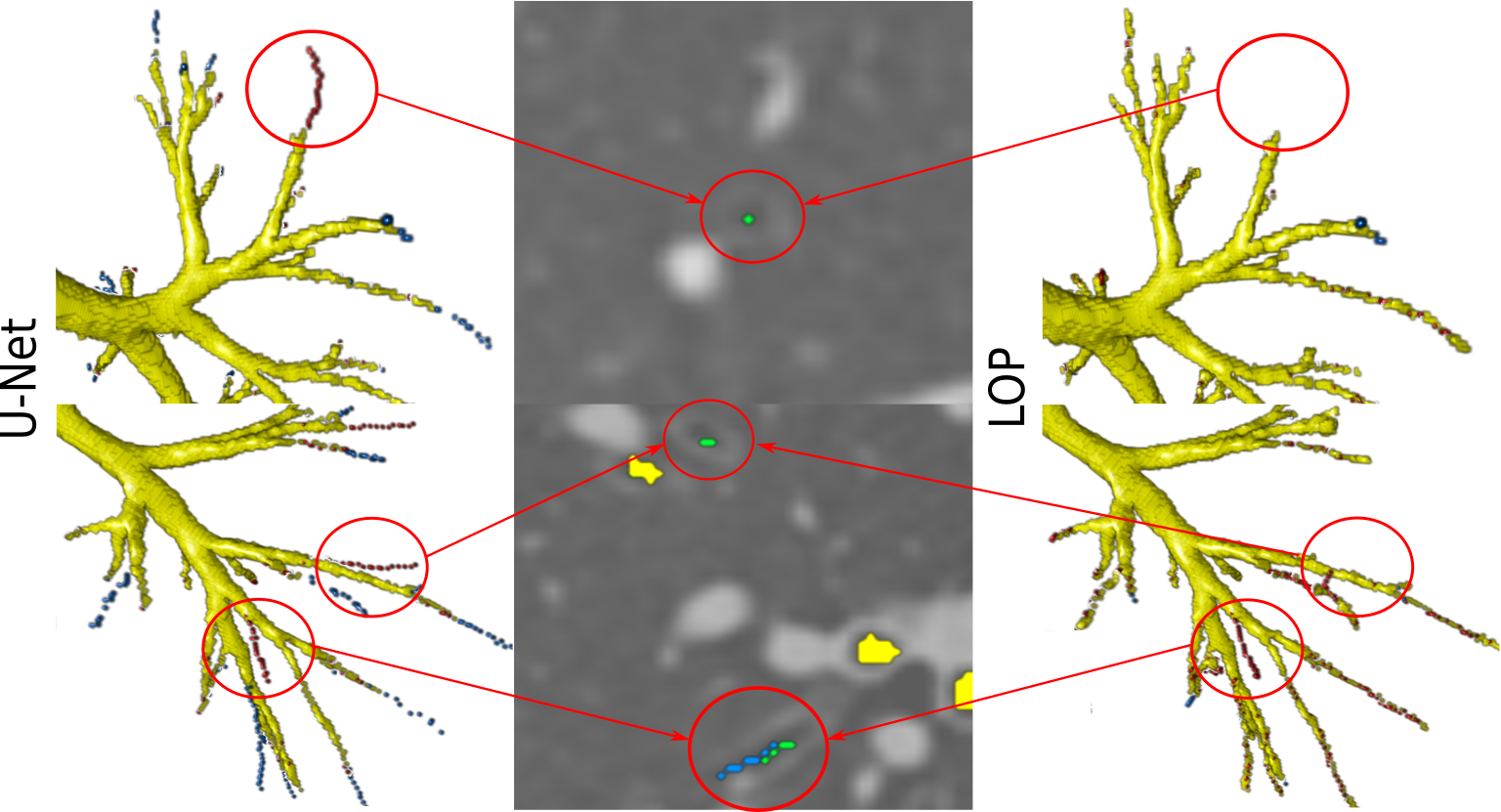}
\caption{Visualization of examples of predicted false positive branches on the DLCST data, obtained by our U-Net-based method and missed by the kNN-VS and LOP methods, that are actually real airways missing in the reference segmentations. The last earmarked branch is detected by the LOP method but with errors: with leakage outside the airway wall. In the 3D visualizations the true positives are displayed in yellow, centerline false negatives in blue and centerline false positives in red. We also show the overlay of the predicted airway centerline masks with the CT scan, showing that the earmarked branches are real airways. The predicted centerlines by our U-Net method are displayed in green, those by the LOP method in blue, and the reference segmentations in yellow.}
\label{figFPrealBranchDLCST}
\end{figure}

Through visual inspection of the results, we observed that several false positive errors correspond to real airways on the CT scan that were missing in the reference segmentations. This is because this reference was built in a conservative way from automatic airway extractions, where branches not extracted by either method~\cite{Lo2010,Lo2009} were not included, and branches evaluated as "partly wrong" were removed, as explained in Section~\ref{sec:GenerationGroundtruth}. Thus, the reference segmentations are somewhat incomplete. For the segmentation results obtained by the three methods, we inspected any false positives that were of tubular shape and that were long enough to be clearly perceived as possible airways. We found 41 such cases of false positives that were actual airways when they were overlaid and analyzed on the CT scan. Out of these 41 misclassified branches, 20 were segmented only by our U-Net-based method, 14 were segmented by our U-Net-based method and either the kNN-VS or LOP method, and 7 were segmented by either the kNN-VS or LOP method and not the U-Net-based method. All cases segmented by our U-Net-based method were free of leakage. In contrast, all cases segmented by the kNN-VS and LOP methods had errors: they were branches longer than the real airway, or had leakage outside the airway wall. We show in Fig.~\ref{figFPrealBranchDLCST} some examples of these misclassified false positive branches, where in the corresponding overlay of the airway mask with the CT scan we can see that they are real airways. Because of this, the CL leakage reported for the three methods in Fig.~\ref{figBoxplotResDLCST} are presumably overestimated, and to a larger extent for our U-Net-based method. Interestingly, most of the 34 misclassified but correctly segmented branches by our U-Net-based method were on CT scans from subjects with COPD: 26 out of 34 (76\%).

Comparing the measures from our U-Net-based method between subjects with normal lung function and subjects with COPD, we found no significant differences in TL ($81.0(78.8-81.5)$ compared to $82.7(80.0-85.2)$, $p=0.15$), but they were significant in CL ($6.27(5.25-8.86)$ compared to $9.18(7.64-10.4)$, $p<0.01$) and in total tree length ($190(174-247)$ compared to $131(123-169)$, $p<0.01$). This indicates that our method has slightly more false positives errors on scans from subjects with COPD. However, this is partly explained by the higher number of misclassified but correctly segmented branches that we found in scans with COPD, for the U-Net results. We show these grouped measures in the Appendix.

\subsection{Evaluation on EXACT'09 data}
\label{sec:ResultsEXACT}

\begin{figure}[t]
\centering
\includegraphics[width=0.9\textwidth]{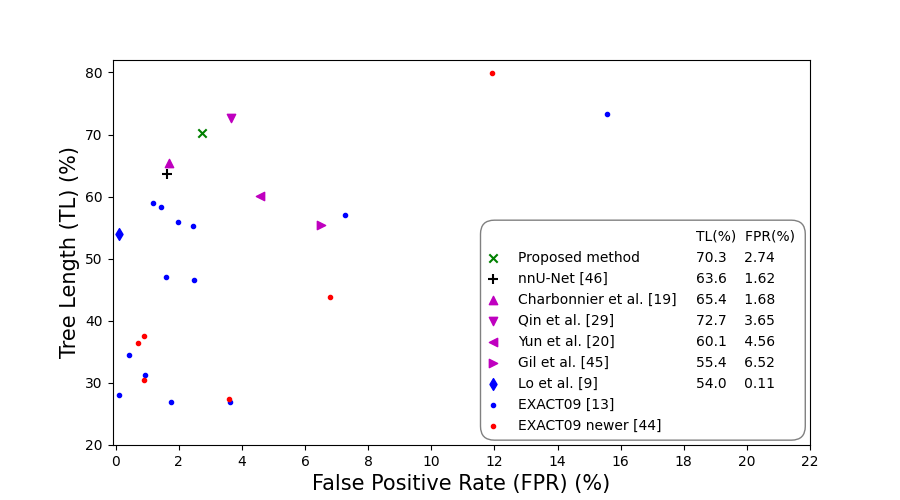}
\caption{Average tree length detected versus average false positive rate over EXACT'09 test set, for our U-Net-based method, the 15 methods from the challenge EXACT'09~\cite{LoExact2012}, the 6 post-challenge methods~\cite{EXACTwebsite}, the 4 recent methods~\cite{Charbonnier2017,Yun2019,Gil2019,Qin2021} evaluated on the EXACT'09 data, and our experiment with the nnU-Net method~\cite{nnUnet2021} on these data. Results are computed by the EXACT'09 challenge organizers except for~\cite{Charbonnier2017}, who did the evaluation slightly differently, leading to slightly better performance, as reported in their paper.}

\label{figResultsEXACT}
\end{figure}

We show in Fig.~\ref{figResultsEXACT} the evaluation on the EXACT'09 data of our U-Net-based method, the 15 methods from the challenge EXACT'09~\cite{LoExact2012}, the 6 post-challenge methods~\cite{EXACTwebsite}, the 4 recent methods~\cite{Charbonnier2017,Yun2019,Gil2019,Qin2021} evaluated on the EXACT'09 data, and our experiment with the nnU-Net method~\cite{nnUnet2021} on these data. The reported results are the average TL and FPR measures obtained over the EXACT'09 test set. Our U-Net-based method achieves an overall TL of $70.3\%$ and FPR of $2.74\%$. When compared to the EXACT'09 methods and the 6 post-challenge methods~\cite{EXACTwebsite}, our U-Net-based method has a TL much higher (at least $10\%$) than the scores from all methods except two, while the reported FPR is only slightly higher ($\approx1\%$) than the average score among these methods. This indicates that our U-Net-based method predicts more complete airway trees on average, and with limited false positive errors. The two methods with higher TL also show a much higher FPR. When compared to the CNN-Leak method~\cite{Charbonnier2017} and the nnU-Net method~\cite{nnUnet2021}, our U-Net-based method has higher completeness but also more false positive errors. The nnU-Net method~\cite{nnUnet2021} achieved a similar performance to the CNN-Leak method~\cite{Charbonnier2017}, with slightly lower TL for a similar FPR score. When compared to the U-Net-FRAD method~\cite{Qin2021}, our U-Net-based method has slightly lower completeness but also less false positive errors. Our method together with the CNN-Leak~\cite{Charbonnier2017}, U-Net-FRAD~\cite{Qin2021} and nnU-Net~\cite{nnUnet2021} methods seem to have the best trade-off between the TL and FPR scores among all tested methods in Fig.~\ref{figResultsEXACT}. However, the authors from~\cite{Charbonnier2017} did not follow the independent evaluation for the EXACT'09 benchmark, but they re-implemented the evaluation algorithm using the reference standard from EXACT'09~\cite{LoExact2012}. Their evaluation on one of the 15 submissions to EXACT'09 resulted in a slightly higher tree length than the one originally reported in~\cite{LoExact2012}. Thus, the TL score reported for this method could be overestimated. Comparing our U-Net with the nnU-Net results, the former shows higher TL ($68.8(61.2-79.7)$ compared to $63.9(52.8-75.2)$, $p=0.01$) and higher FPR ($1.90(0.51-3.86)$ compared to $1.07(0.30-2.25)$, $p=0.04$), but they are not significantly different. This indicates that both our U-Net-based method and the nnU-Net have comparable accuracy. When compared to the 2.5D CNN method~\cite{Yun2019} and the PICASSO method~\cite{Gil2019}, our U-Net-based method shows better performance measures, with higher completeness and less false positive errors.

\begin{figure}[t]
\centering
\includegraphics[width=0.95\textwidth]{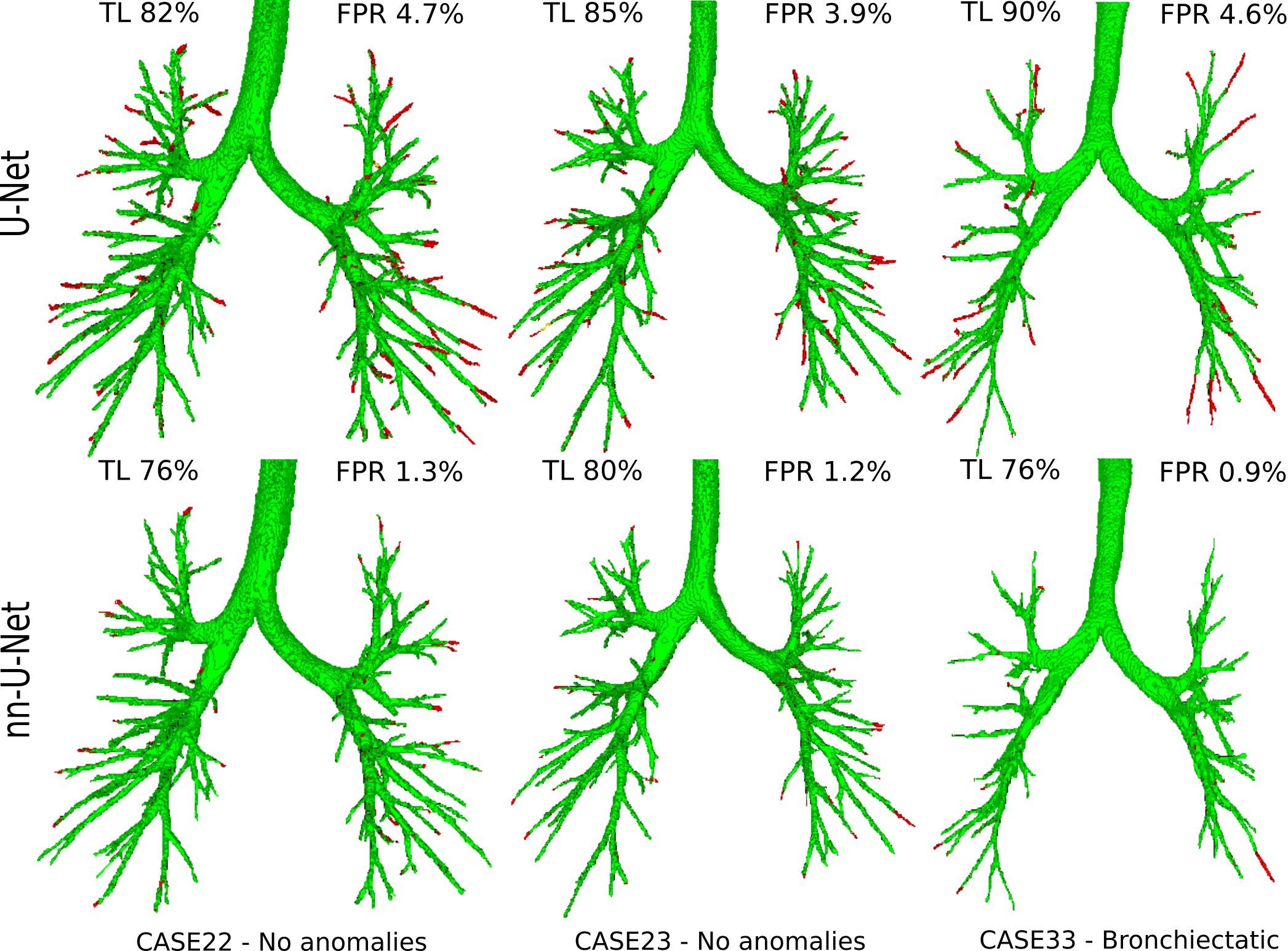}
\caption{Visualization of predicted airway trees on the EXACT'09 data obtained by our U-Net-based method and the nnU-Net method~\cite{nnUnet2021}. True positives are displayed in green and false positives in red, indicated according to the EXACT'09 evaluation. The cases displayed correspond to 3 out of the 5 U-Net results with the highest false positive rate reported in the EXACT'09 evaluation, where we found through visual inspection that many of the reported false positives are real airways on the CT scan, which were presumably missing in the EXACT'09 reference.}
\label{figVisualResEXACT}
\end{figure}

Through visual inspection of the results, we observed that for some CT scans a large number of airways predicted by our U-Net-based method that were reported as false positives in the EXACT'09 evaluation are real airways on the CT scan. We show in Fig.~\ref{figVisualResEXACT} visualizations of airway trees, obtained by our method and the nnU-Net method~\cite{nnUnet2021}, for which most of the reported false positive errors are real airways. Our method shows more misclassified false positive branches than the nnU-Net method~\cite{nnUnet2021}, which partly explain the higher FPR scores in the displayed cases. This is because the reference standard from EXACT'09, constructed from all 15 participating methods to the challenge, is somewhat incomplete: it was estimated that on average approximately $10\%$ more branches were visible in the scans than those that were included in the reference~\cite{LoExact2012}. Thus, the false positive rate reported by our U-Net-based method and the nnU-Net method~\cite{nnUnet2021} in Fig.~\ref{figResultsEXACT} may be overestimated, with a larger extent for our method. This could be also for the other methods~\cite{Charbonnier2017,Yun2019,Gil2019,Qin2021,EXACTwebsite} we compared with, which used the same reference for evaluation.

We compared the measures from our U-Net-based method between a group of scans without any reported anomalies on the CT scan (cases 21, 22, 23, 28, 29, 35, and 37), and a group of scans with reported bronchiectasis (cases 33, 34, 36, and 39)~\cite{LoExact2012}. We found no significant differences in TL ($80.1(65.6-83.2)$ compared to $79.4(75.7-82.1)$, $p=0.59$), in FPR ($2.98(0.99-3.87)$ compared to $3.34(1.99-6.50)$, $p=0.21$) and in total tree length ($116(100-193)$ compared to $275(232-295)$, $p=0.07$). This indicates that our method has similar accuracy with respect to the reference segmentations on both scans without anomalies and scans with bronchiectasis. We show these grouped measures in the Appendix.


\section{Discussion}
\label{sec:Discussion}

\subsection{Performance compared to other methods}
The proposed U-Net-based method has shown good performance on the three different datasets, achieving highly complete airway segmentations with low false positive errors. The three datasets tested have very different characteristics: with subjects of a wide age range (from children in CF-CT data to older adults in DLCST data) and with airway abnormalities from diverse lung diseases, including CF and COPD. On the CF-CT and DLCST data, our U-Net-based method obtained more accurate airway segmentations than the other tested methods~\cite{PerezRovira2016,Lo2010,Lo2009}, shown by higher completeness and Dice scores altogether, and lower false positive errors. On the EXACT'09 data, our U-Net-based method achieved performance measures similar to the best performing previous methods, with the second highest sensitivity score among all methods that reported good specificity. Especially, it is noticeable that our method achieved this good performance on the highly heterogeneous EXACT'09 data with models trained on different and more homogeneous datasets (CF-CT and DLCST). This shows the robustness and the capacity to generalize across different data of the proposed method. Furthermore, our U-Net-based method obtained similar performance measures on scans from both healthy and diseased subjects, on the CF-CT and EXACT'09 data. On the DLCST data, our method had slightly more false positives errors on scans from subjects with COPD, where airway detection can be more challenging due to emphysema. However, on scans with COPD we reported a higher number of false positives in the U-Net results that were real airways on the CT scan, which partly explains the lower specificity in the diseased cases. On the CF-CT data, the method segmented more airway branches on scans from subjects with CF, probably due to the widening of peripheral airways due to CF-bronchiectasis, which makes them easier to detect on the CT scan.

When compared to other CNN-based methods evaluated on the EXACT'09 data, our U-Net-based method shows a performance similar to that of the CNN-Leak~\cite{Charbonnier2017}, U-Net-FRAD~\cite{Qin2021} and nnU-Net~\cite{nnUnet2021} methods. These four methods seem to have the best trade-off between sensitivity and specificity among all tested methods in Fig.~\ref{figResultsEXACT}. However, the U-Net-FRAD method~\cite{Qin2021} was trained on data that included the EXACT'09 training set, using their own manually-drawn reference segmentations on these data. Since the scans from the EXACT'09 training and test sets have similar characteristics and scanning parameters~\cite{LoExact2012}, this gives an advantage to the U-Net-FRAD method~\cite{Qin2021} over our U-Net-based method, which was trained on different data. On the other hand, the CNN-Leak method~\cite{Charbonnier2017} requires a coarse airway segmentation as initialization to the CNN-based leakage removal algorithm, and thus the completeness of the results is limited by that of the initial segmentation. In contrast, our U-Net-based method segments the full airway tree directly. Moreover, CNN-Leak~\cite{Charbonnier2017} applied leakage removal to a series of 15 runs of the region growing based algorithm~\cite{Rikxoort2009}, with different parameters that control the extent of leakage, upon which the results were merged. In fact, the results from applying the leakage removal only to the baseline segmentation~\cite{Rikxoort2009} had a much lower sensitivity (TL $51.8\%$ in contrast to $65.4\%$) and a slightly higher specificity (FPR $1.01\%$ in contrast to $1.68\%$). Similarly, an ensemble of U-Net results with different settings would likely have a slightly better performance, but this is more time consuming. When compared with the nnU-Net method~\cite{nnUnet2021}, our U-Net-based method achieved slightly higher completeness and false positive errors, but we found that the differences were not statistically significant. When compared with the 2.5D CNN method~\cite{Yun2019}, our U-Net-based method shows better performance with both higher sensitivity and specificity. The higher accuracy of our method can be because the 2.5D CNN method processes three perpendicular 2D slices around each voxel, while the 3D U-Net can better capture the 3D morphological information of airways.

Regarding computational efficiency, our U-Net-based method has a low execution time during inference of about 1 min per scan, including all post-processing steps to obtain a single connected binary airway tree. It should be noted that computation times of different methods cannot easily be compared due to differences in hardware, but the following may give a rough idea of the efficiency of our method compared to previous work. The execution times reported by other methods evaluated on the EXACT'09 data are: the kNN-VS method~\cite{Lo2010}, 55 min per scan, although it was run on a much older CPU and was not parallelized; the CNN-Leak method~\cite{Charbonnier2017}, 3-5 min per scan; the U-Net-FRAD method~\cite{Qin2021}, approximately 50 s per scan, although it does not include any post-processing steps; the 2.5D CNN method~\cite{Yun2019}, 2-8 min per scan; and the PICASSO method~\cite{Gil2019}, approximately 10 min per scan. Our experiments with the nnU-Net method~\cite{nnUnet2021} required 5-15 min per scan, although this includes some expensive pre-processing operations applied on the fly to the test images, such as resampling. Additionally, the U-Net method~\cite{Nadeem2021} reports approximately 6.5 min per scan.

\subsection{Advantages of the proposed method}
\label{sec:AdvantagesMethod}

The proposed U-Net-based method processes large 3D patches extracted from the CT scan in a single pass through the network. This makes our method simple, robust, and efficient at inference time, as only a few large patches are processed to segment a full scan. Although processing large input patches does not necessarily imply better performance, we observed that using large input patches resulted in better results and faster convergence of the training and validation losses. This is likely due to the more efficient use of the data during training. In contrast, when using smaller patches to train the model, this requires sampling patches at every voxel to equally process the same region of the CT scan comprised in a larger patch, which results in a high number of total patches. Moreover, since convolution operations are accelerated on the GPU, it is more efficient computationally to process a large patch at once through the network, rather than sequentially loading to GPU and processing smaller patches to include the same large region of the CT. In contrast to our approach, other U-Net-based methods in the literature either i) apply a small U-Net locally around detected candidate airway locations, processing many image patches per scan~\cite{Jin2017,Meng2017}; or ii) apply a model with larger memory footprint on smaller image patches in a sliding-window fashion~\cite{Qin2019,Qin2019-2,Zhao2019,Wang2019,Qin2020,Qin2021,Zheng2020,Zhou2021,Nadeem2021}. With our simple U-Net-based method and without further processing of the network output (except for computing the largest connected component) we obtained highly complete and accurate airway segmentations on the three datasets tested. On the EXACT'09 data, our method achieved a similar performance to the more complex U-Net-FRAD method~\cite{Qin2021}, which used manually annotated EXACT'09 data for model training; and the nnU-Net method~\cite{nnUnet2021}, using a similar U-Net as backbone. An additional advantage of processing few, large patches by our method is that fewer edge artifacts are introduced when tiling together the predicted output patches of the network. These artifacts typically occur where the tiled predicted patches meet in the full-size output (or where the patches overlap if this occurs), because the predictions from each patch can be slightly different due to border effects when using non-valid convolutions, and can cause discontinuities in the predicted airway mask.

\subsection{Limitations}
\label{sec:LimitationsMethod}

A limitation of our validation of the proposed method is that we did not evaluate it on CT scans with severe airway disease: the CF-CT dataset include subjects with moderate CF disease, the DLCST dataset include subjects with moderate COPD, and the EXACT'09 test data has various airways abnormalities but only one scan with reported "extensive bronchiectasis"~\cite{LoExact2012}. Testing the method on severe cases would be important to assess its generalizability to tackle challenging airways with abnormally deformed shapes due to severe disease. Moreover, our evaluation of segmentation performance with respect to the presence of disease on the DLCST dataset may not show the whole picture. This is because the reference segmentations on these data were built in a conservative way from automatic airway extractions~\cite{Lo2010,Lo2009}, and could be less complete for scans with severe emphysema.

A limitation of our U-Net-based method is that the prediction of the airway tree output by the U-Net is not guaranteed to form a connected tree structure.  This could complicate the automatic extraction of airway biomarkers based on these segmentations as some methods assume a fully connected tree as input~\cite{Petersen2014-2,PerezRovira2016,Kuo2020}. The airway predictions obtained for this paper had typically some segmented peripheral airways disconnected from the main airway tree, and these were discarded when computing the largest connected component, which reduced the completeness of our airway segmentations. Alternatively, airway measurements techniques that do not rely on fully connected trees, such as~\cite{Estepar2012}, could be used. Several methods have been proposed to extract more complete, connected tree structures. The voxel-connectivity U-Net formulation~\cite{Qin2019} aims to improve connectivity in the airway prediction, resulting however in a model that is significantly more complex than our U-Net used. The linear programming-based tracking module on top of U-Net~\cite{Zhao2019} attempts to link disconnected components of airways from the U-Net output. The mean-field networks and graph neural networks (GNNs)~\cite{Selvan2020} emphasize the prediction of connected tree-like structures by phrasing the tree extraction problem as graph refinement starting from an over-connected input graph. The joint U-Net-GNN method~\cite{GarciaUceda2019} attempts to integrate this tree-like modelling in the U-Net prediction. However, none of these methods can guarantee fully connected airway tree predictions.

Finally, the memory footprint of our U-Net-based method could be further reduced, which would allow us to fit even larger images to the network, and possibly the entire CT scan. It may be possible to reduce the number of feature maps, especially in the decoding path of the U-Net, without decreasing much the performance. Also, using the partially reversible U-Net formulation\cite{Brugger2019} in our method could largely reduce its memory footprint, by lowering the number of activation maps in the network stored in memory. However, this may result in an increase of training time (the authors from~\cite{Brugger2019} reported a 50\% increase for their tested 5-level U-Net model, similar to our U-Net).


\section{Conclusions}
\label{sec:Conclusions}
In this paper, we present a fully automatic and end-to-end optimised method to segment the airways from thoracic CT, based on the U-Net architecture. In contrast to previous U-Net methods for airway segmentation, the proposed method processes large 3D image patches often covering entire lungs. This is achieved by using a simple and low-memory 3D U-Net as backbone. This makes the method robust and efficient, which is important if the method is deployed in clinical software. Our method obtained highly complete and accurate airway segmentations on three very different datasets including CT scans with various airway and lung abnormalities. On the EXACT'09 test set, our method achieved the second highest sensitivity score among all methods that reported good specificity; and it outperformed previous methods on the other datasets.


\bibliography{refers.bib}


\section*{Acknowledgements}

The research leading to these results has received support from the Innovative Medicines Initiative Joint Undertaking under grant agreement n. 115721 resources of which are composed of financial contribution from the European Union's Seventh Framework Programme (FP7/2007-2013) and EFPIA companies’ in kind contribution.


\section*{Author contributions}

A.G. did the literature research, conducted the experiments, analysed the results, and wrote the manuscript; R.S. contributed to the analysis of results and writing of the manuscript; Z.S. collected data, H.T. collected data, M.B. supervised the literature research, experiments and analysis of results, and contributed to writing of the manuscript. All authors reviewed the manuscript.


\section*{Additional information}

\subsection*{Code availability}
The source code for our implementation of the proposed method, and the trained model for the results obtained on the EXACT'09 test set, are available here: \url{https://github.com/antonioguj/bronchinet}

\subsection*{Competing interests}
The author(s) declare no competing interests.


\section*{Appendix}
\subsection*{Learning curves of the proposed method}
\label{sec:learningCurve}

\begin{figure}[ht]
\centering
\begin{subfigure}[t]{0.32\textwidth}
\includegraphics[width=\textwidth]{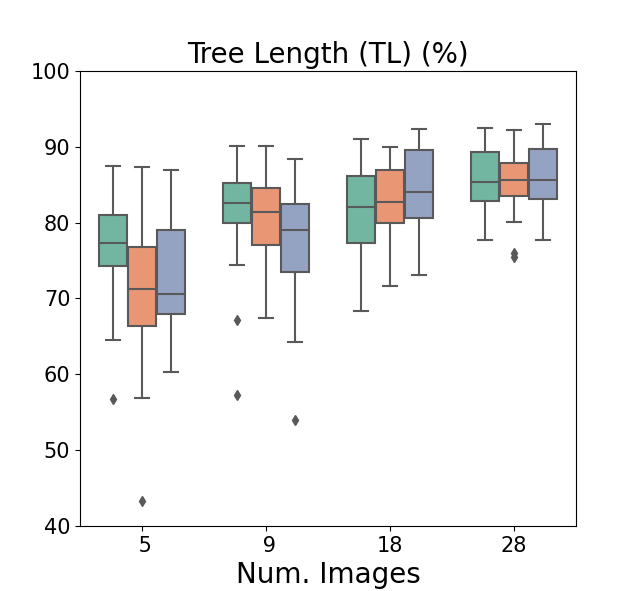}
\end{subfigure}
\begin{subfigure}[t]{0.32\textwidth}
\includegraphics[width=\textwidth]{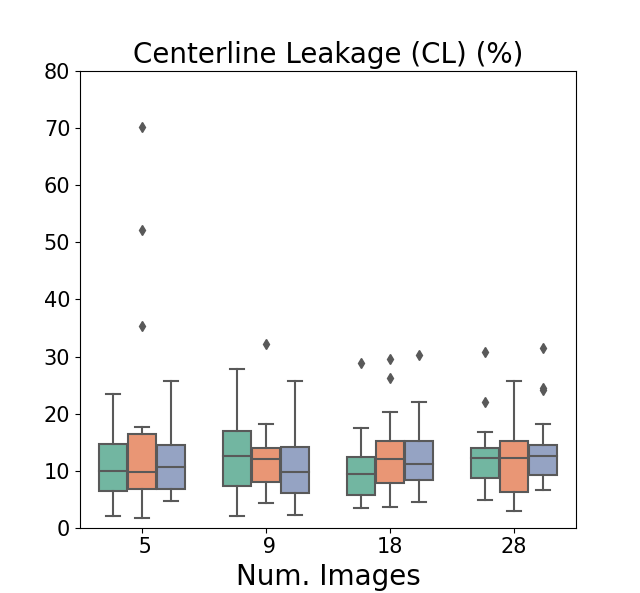}
\end{subfigure}
\begin{subfigure}[t]{0.32\textwidth}
\includegraphics[width=\textwidth]{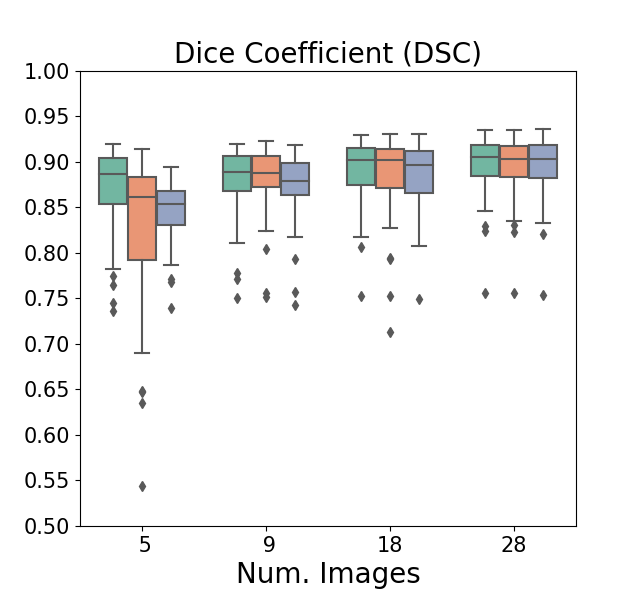}
\end{subfigure}
\caption{Learning curve for our U-Net-based-method trained on both CF-CT and DLCST data together. Boxplots showing i) tree length detected, ii) centerline leakage and iii) Dice coefficient on both CF-CT and DLCST data together, for each experiment and training size. For each boxplot, the box shows the quartiles of the data (defined by the median, 25\% and 75\% percentiles), the whiskers extend to include the data within 1.5 times the interquartile range from the box limits, and the markers show the data outliers.}
\label{figLearnCurve}
\end{figure}

We computed the learning curve for the proposed U-Net-based-method trained on both CF-CT and DLCST data together. To do this, we trained several models with different sizes of the training data. The maximum training size has half of the CT scans from the CF-CT and DLCST data (28 scans in total), which is the same data we used to train the model evaluated on the EXACT'09 data in this paper. We keep the remaining 28 scans for testing the trained models. For each training set used, the ratio between CF-CT and DLCST scans is the same as in the full dataset. We did three experiments for each training size, with randomly assigned training images (except for the largest run with 28 scans). To compute the airway predictions on the test data, we did not extract the largest connected component from the thresholded output of the U-Net, as we did for the other experiments in this paper. This is to account for the full prediction of the U-Net in assessing the method accuracy for all training sizes. To compare the results for different training sizes, we applied the paired, two-sided Student's T-test on the average of the measures from the three experiments for a given size, and consider that a p-value lower than 0.01 indicates a significant difference. We show in Fig.~\ref{figLearnCurve} the computed learning curves, with the different performance metrics obtained for each run and training set size. The measures of tree length detected increase progressively with the training size. The difference between the scores with sizes of 18 and 28 images is still significant ($p<0.001$), and adding more training images could still improve slightly the results. For the measures of centerline leakage and Dice coefficient, they are more similar between sizes of 9 and 18 images ($p=0.35$ and $p=0.26$, respectively) and between sizes of 18 and 28 images ($p=0.99$ and $p=0.019$, respectively).

\newpage
\subsection*{Results of the proposed method grouped by the presence of lung disease}
\label{sec:resultDisease}

\begin{figure}[ht]
\centering
\begin{subfigure}[t]{0.32\textwidth}
\includegraphics[width=\textwidth]{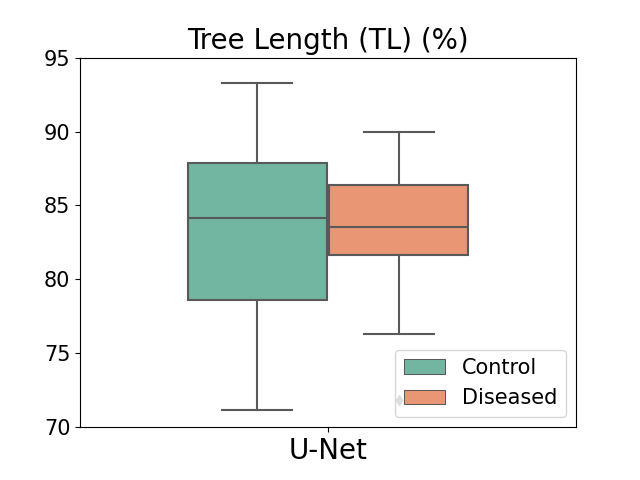}
\end{subfigure}
\begin{subfigure}[t]{0.32\textwidth}
\includegraphics[width=\textwidth]{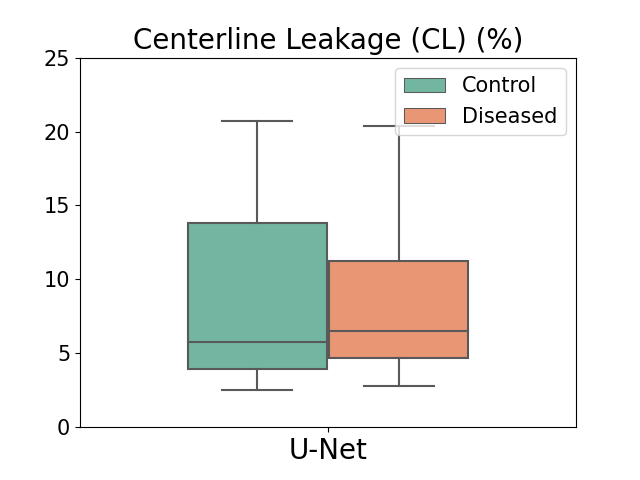}
\end{subfigure}
\begin{subfigure}[t]{0.32\textwidth}
\includegraphics[width=\textwidth]{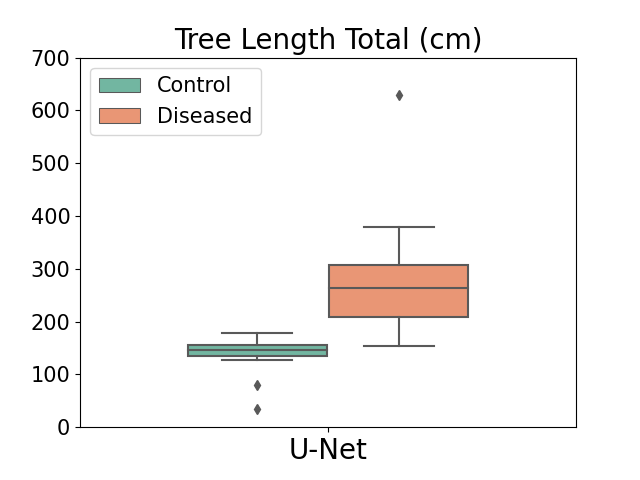}
\end{subfigure}
\caption{Boxplots showing i) tree length detected, ii) centerline leakage and iii) total tree length on the CF-CT data, grouped by the presence of CF disease in the CT scans, for the results obtained with our U-Net-based method. For each boxplot, the box shows the quartiles of the data (defined by the median, 25\% and 75\% percentiles), the whiskers extend to include the data within 1.5 times the interquartile range from the box limits, and the markers show the data outliers.}
\label{figBoxplotResDiseaseCFCT}
\end{figure}

\begin{figure}[ht]
\centering
\begin{subfigure}[t]{0.32\textwidth}
\includegraphics[width=\textwidth]{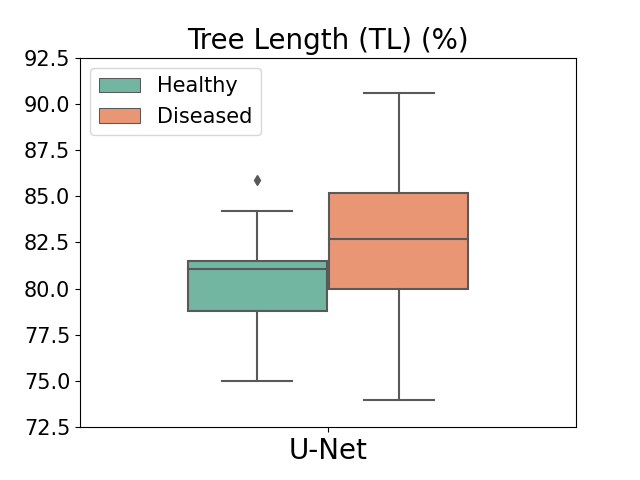}
\end{subfigure}
\begin{subfigure}[t]{0.32\textwidth}
\includegraphics[width=\textwidth]{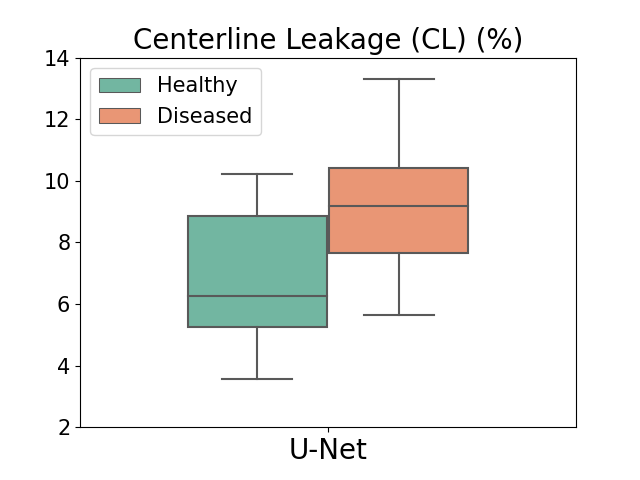}
\end{subfigure}
\begin{subfigure}[t]{0.32\textwidth}
\includegraphics[width=\textwidth]{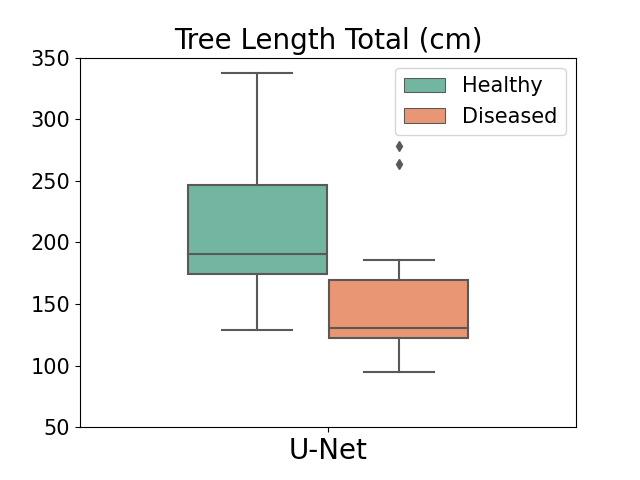}
\end{subfigure}
\caption{Boxplots showing i) tree length detected, ii) centerline leakage and iii) total tree length on the DLCST data, grouped by the presence of COPD disease in the CT scans, for the results obtained with our U-Net-based method. For each boxplot, the box shows the quartiles of the data (defined by the median, 25\% and 75\% percentiles), the whiskers extend to include the data within 1.5 times the interquartile range from the box limits, and the markers show the data outliers.}
\label{figBoxplotResDiseaseDLCST}
\end{figure}

\begin{figure}[ht!]
\centering
\begin{subfigure}[t]{0.32\textwidth}
\includegraphics[width=\textwidth]{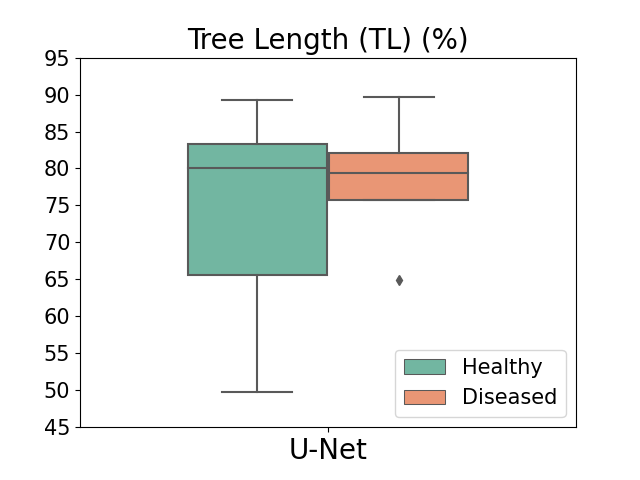}
\end{subfigure}
\begin{subfigure}[t]{0.32\textwidth}
\includegraphics[width=\textwidth]{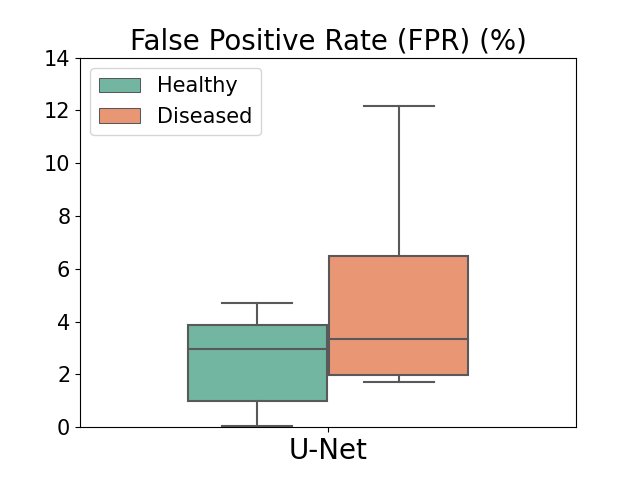}
\end{subfigure}
\begin{subfigure}[t]{0.32\textwidth}
\includegraphics[width=\textwidth]{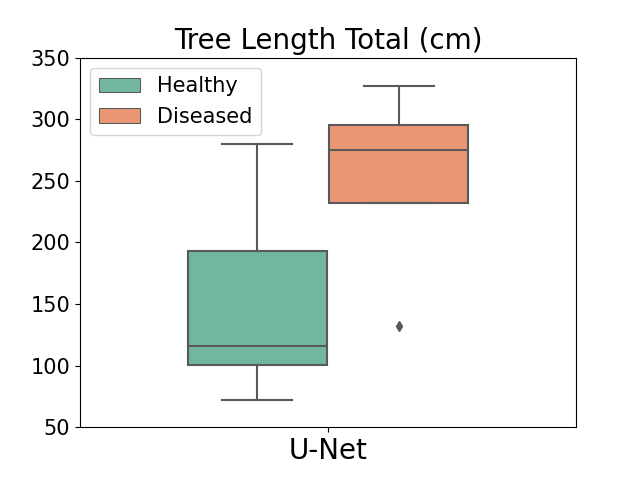}
\end{subfigure}
\caption{Boxplots showing i) tree length detected, ii) false positive rate and iii) total tree length on the EXACT'09 test data, grouped by the presence of bronchiectasis disease in the CT scans, for the results obtained with our U-Net-based method. The healthy group contains 7 scans without any reported anomalies on the CT scan, while the diseased group contains 4 scans with reported bronchiectasis [13]. For each boxplot, the box shows the quartiles of the data (defined by the median, 25\% and 75\% percentiles), the whiskers extend to include the data within 1.5 times the interquartile range from the box limits, and the markers show the data outliers.}
\label{figBoxplotResDiseaseEXACT}
\end{figure}
\pagebreak
\subsection*{Implementation details of the nnU-Net method and our experiments for airway segmentation}
\label{sec:experimentNNUnet}

The nnU-Net method proposed by Isensee et al.~\cite{nnUnet2021} is a general segmentation framework designed for biomedical segmentation tasks. We applied the method for airway segmentation from chest CTs. We used the implementation in~\url{https://github.com/MIC-DKFZ/nnUNet}.

For our experiment, we used the so-called "full3d\_UNet" in the nnU-Net framework, which is the most similar to the U-Net in our method. This U-Net has 5 levels of resolution, where each level in the downsampling / upsampling path has two $3\timestxt3\timestxt3$ convolutional layers and a $2\timestxt2\timestxt2$ pooling / deconvolution layer, respectively. Each convolution operation is followed by an instance normalisation layer and a leaky rectified linear (leaky-ReLU) activation. The number of features in the top resolution level is 32, and after each pooling / deconvolution layer the number of feature channels is halved / doubled, respectively. The network uses deep supervision at all levels of the U-Net. In this, the output of every last convolutional layer at every resolution level is concatenated, after being resampled to the original resolution. Then, the final layer is a 1x1x1 convolutional layer followed by a sigmoid activation function.

To train the network, we used the same 28 CT scans and ground truth segmentations from the CF-CT and DLCST data as we used to train our method. The nnU-Net method uses as training loss function a combination of the i) binary cross entropy and ii) soft Dice losses. We could not modify the loss computation in the nnU-Net to consider only voxels within the lung regions, as we did for our method in equation (1). Instead, we masked the ground truth segmentations to the mask of the lung fields, to remove the trachea and part of the main bronchi. The nnU-Net method uses the SGD optimizer with an adaptable learning rate, starting with a value of $10^{-2}$. We trained the model for a sufficiently large number of epochs, 600, until the training and validation losses are clearly stabilized. We then retrieved the model with the overall minimum validation loss for testing, denoted by "model\_best" in the nnU-Net framework. Training time was approximately 2-3 days on a GPU GeForce RTX 2080 Ti. Test time inference takes between 5-15 min per scan, including pre-processing.

The nnU-Net applies some pre-processing operations to the scans and ground truth in the training dataset. First, the images are cropped to the region of non-zero values in the ground truth airway masks. Then, the images are resampled to a fixed resolution equal to the median over the training dataset, using 3rd order spline interpolation for the CTs and nearest neighbour interpolation for the ground truth masks. During training, the nnU-Net extracts random  patches from the scans and ground truth, with a total of 1 patch per image and per epoch. Then, random rigid transformations are applied for data augmentation, including i) flipping in the three directions, ii) random small 3D rotations, and iii) random scaling.

At inference time, the nnU-Net applies the same pre-processing operations to the test scans as those applied on the training data used for the tested model. The images are resampled to the same fixed resolution used for the training data, i.e. the median over these data. Input patches to the network are extracted from the scans in a sliding-window fashion, with an overlap of roughly $50\%$ between the patches in each direction, and then are processed through the trained model. The output predicted patches containing airway probability maps are aggregated and the full-size output is reconstructed. Then, thresholding is used to obtain the airway binary segmentation, and the pre-processing steps are reversed to recover the original image resolution. Finally, we merge this segmentation with a mask for the trachea, main bronchi, and the first 5 voxels of the next branches to obtain the full airway tree, which is easily computed by a region growing method~\cite{Lo2010}.

\end{document}